# Half-metallicity and anisotropy magnetoresistance properties of Heusler alloys $Fe_2Co_{1-x}Cr_xSi$

Y. Du[1], G. Z. Xu[1], E. K. Liu[1], G. J. Li[1], H.G. Zhang[1], S. Y. Yu[2], W. H. Wang, [1,a)] and G. H. Wu[1]

[1] *Beijing National Laboratory for Condensed Matter Physics, Institute of Physics, Chinese Academy of Sciences, Beijing 100080, P. R. China*

[2] *School of Physics, Shandong University, Jinan 250100, China*

**Abstract:**

In this paper, we investigate the half-metallicity of Heusler alloys $Fe_2Co_{1-x}Cr_xSi$ by first principles calculations and anisotropy magnetoresistance measurements. It is found that, with the increase of Cr content x, the Fermi level of $Fe_2Co_{1-x}Cr_xSi$ moves from the top of valence band to the bottom of conduction band, and a large half-metallic band gap of 0.75 eV is obtained for x=0.75. We then successfully synthesized a series Heusler $Fe_2Co_{1-x}Cr_xSi$ polycrystalline ribbon samples. The results of X-ray diffraction indicate that the $Fe_2Co_{1-x}Cr_xSi$ series of samples are pure phase with a high degree of order and the saturation magnetic moment follows half-metallic Slater-Pauling rule. Except for the two end members, $Fe_2CoSi$ and $Fe_2CrSi$, the anisotropic magnetoresistance of $Fe_2Co_{1-x}Cr_xSi$ (x=0.25, 0.5, 0.75) show a negative value suggesting they are stable half-metallic ferromagnets.

## 1.Introduction

Half-metallic ferromagnets were discovered by theory with the prediction of 100% spin polarization [1,2], in which the majority-spin electrons are metallic, whereas the minority-spin electrons are semiconducting. Since the first prediction of half-metallic ferromagnet in Heusler compounds in 1983 by de. Groot [3], several Heusler series that possess half-metallic properties have been experimentally realized [4-6]. Full-Heusler compounds are usually presented as general formula $X_2YZ$, where X and Y are transition elements, Z represents main group element. The cubic unit cell consists of four interpenetrating face-centered-cubic (fcc) sublattices, two of which are occupied by identical X atoms and the other two by Y and Z atoms, respectively [7]. Because of high Curie temperature, Co-based Heusler compounds are considered as the most promising candidates for half-metallic ferromagnets working at room temperature [8,9]. However, Co-based Heusler compounds have a serious inadequacy that a high $L2_1$ degree of order is necessary to present a high spin polarization. For example, the recent experiments found that, in sputtered $Co_2MnSi$ Heusler films, a few degree of atomic disorder between Co and Mn will cause a largely reducing of the spin polarization [10]. On the other hand, the thermal excitation and spin-flip scattering can also destroy the half-metallitity in Co-based Heusler alloys. To avoid thermal excitation and spin-flip scattering of electrons to the conduction sub-band, Benjamin Bakle *et al.* [11] proposed a method for engineering the band gap and tuning the Fermi level by doping Fe in Heusler alloys $Co_2Mn_{1-x}Fe_xSi$. Very recently, based on the first principles calculations, Luo *et al.* [12] found that Fe-based Heusler alloys $Fe_2CoSi$ and $Fe_2CrSi$ are two typical half-metallic ferromagnets. They also pointed out that the half-metallitity of these Fe-based Heusler alloys is insensitive to atomic disorder. However, the Fermi level of $Fe_2CoSi$ and $Fe_2CrSi$ locate at the edge of valance and conductance bands, respectively. In this paper, using the first principles calculations, we investigate the band structures of a series $Fe_2Co_{1-x}Cr_xSi$ Heusler compounds, to find the most stable half-metallic ferromagnets. The in-plane anisotropic magnetoresitance (AMR) experiments were further performed to verify the half-metallicity in this series.

## 2. Experimental

$Fe_2Co_{1-x}Cr_xSi$ (x=0.0, 0.25, 0.5, 0.75, 1) ingots were prepared by arc-melting under an Ar atmosphere. Melting was repeated several times to obtain chemically homogenous ingots. Because the flaked sample for the measurements of AMR cannot be obtained by arc-melting, the melt-spinning method was utilized to make small metal sheets from the obtained ingots. The thickness of the ribbon samples used in this work is roughly around 46 μm. Structural examination was measured by x-ray diffraction (XRD) with Cu- Ka radiation. The AMR effect in all samples was measured by the standard four-terminal method in Physical Property Measurement System (PPMS). A superconducting quantum interference device (SQUID) was used to determine the saturation magnetic moment in the low temperature of 5K. The calculations of density of states (DOS) and Band structure have been performed on basis of the density-functional theory (DFT) within the general gradient approximation (GGA). We also used GGA+U calculations to account for the on-site electron-electron Coulomb interaction.

## 3. Results and discussion

Figure 1 shows the XRD patterns for the $Fe_2Co_{1-x}Cr_xSi$ with different Cr content x measured at room temperature. Indexing the characteristic reflections, we find that the studied samples crystallized in face centered cubic (F.C.C) structure without other detectable second phases. The left part of Fig. 1 shows the XRD superlattice reflections measured by the step-scan method. The appearance of superlattice (111) and (200) indicates that our samples have a highly ordered structure [13,14], and the values of the degree of atomic ordering can be roughly estimated from the ratio of the integrated intensity of the superlattice reflection (111) and (200). It is well known that, the intensities of the *B*2 and *L*$2_1$ (or $Hg_2CuTi$) superlattice reflection are proportional to $S^2$ and$(1-2\alpha)^2S^2$, respectively. As we have mentioned above, the chemical formula of full-Heusler alloys can be presented as $X_2YZ$ (*X*: Fe, *Y*: Cr，Co, *Z*: Si). The S parameter represents the *B*2 order from the *A2* structure, and S=1 indicates the alloys has a perfectly *B*2-ordered structure. On the other hand, in the *A2* structure, both the X and Y sites are randomly occupied by X and Y atoms, thus S should be zero. In addition, α indicates the ordering from *B*2 to *L*$2_1$. As a result, we can obtain the S and α values from the formulas as shown below[13],

$$S = ((I_{200}/I_{220})_{experiment}/(I_{200}/I_{220})_{theory})^{1/2} \quad (1)$$

$$(1-2\alpha)S = ((I_{111}/I_{220})_{experiment}/(I_{111}/I_{220})_{theory})^{1/2} \quad (2)$$

The calculated values of S and α are shown in the inset of Fig. 1. All the studied samples exhibit a almost same S value of 0.9, which indicates nearly 5% of the X site is occupied by improper atoms (Y and Z) (A2-disorder). On the other hand, the values of α are very small, showing a local maximum of 0.2 at x=0.5, which means about 20% of the Y (or Z) site is occupied by Z (or Y) atoms (*B*2-disorder). We therefore can conclude that all the samples are approximate highly ordered *L*$2_1$ or $Hg_2CuTi$ structure.

Based on the calculations by Luo *et al.* [12], the Fermi level of $Fe_2CrSi$ and $Fe_2CoSi$ lies in the top of valance band and the bottom of conduction band, respectively. Therefore their half-metallitity are easily destroyed by the thermal excitation and/or spin-flip scattering. Following the idea of tuning the Fermi level position in Heusler alloys $Co_2Fe_{1-x}Mn_xSi$ [11] and $Co_2FeAl_{1-x}Si_x$ [14,15], one can expect to adjust the Fermi level by doping Cr in $Fe_2Co_{1-x}Cr_xSi$. In Fig. 2 (a), we show the calculated spin-resolved DOS for $Fe_2Co_{1-x}Cr_xSi$ with different Cr content x. The upper part of each panel displays the majority spin densities and the lower one the minority spin densities. The results of DOS in Heusler alloys $Fe_2CrSi$ and $Fe_2CoSi$ are consistent with the previous results [12]. The majority states of $Fe_2CrSi$ exhibit a rather high density in the vicinity of Fermi level that is caused by weakly dispersing, flat bands. As expected, we successfully shift the position of the Fermi level to the center of minority gap from the edge with the increase of the Cr substitution.

In order to clear show the motion of minority spin band near Fermi level, in Fig. 2(b), we display the energy of top valance band $E_{VB}$ (black circles) and bottom conductance band $E_{CB}$ (red circles) as a function of the Cr composition x. We found that, with the increase of Cr content x, $E_{VB}$ first decreases and reaches a local minimum at x=0.75 and then increases. On the contrary, $E_{CB}$ exhibits opposite tendency. We further calculated the sizes of half-metallic band gap $\Delta E_{gap}$ (= $E_{VB}$-$E_{CB}$), and the results are plotted as a function of x in Fig. 2 (c). A very small $\Delta E_{gap}$ of 0.05eV is found for $Fe_2CoSi$, however, if Co is partially replaced by Cr and the $\Delta E_{gap}$ will rapidly increase with increasing Cr content x, reaching a maximum value of 0.71eV at x=0.75. The enlarged profiles for spin-resolved DOS of $Fe_2CoSi$ and $Fe_2Co_{0.25}Cr_{0.75}Si$ are shown in the inset of Fig.2(c). Interestingly, unlike $Fe_2Co_{0.25}Cr_{0.75}Si$ having a large half-metallic band gap, for $Fe_2CoSi$, there exists only a touching point at the Fermi level in the down-spin channel.

Figure 3 shows the saturation magnetization values ($M_s$) as a function of Cr composition x. The inset shows the typical magnetization of all the samples measured at 5K by a SQUID magnetometer. The $M_s$ values are 4. 88, 4. 14, 3. 58, 2. 78 and 2.1 (in $\mu_B$/formula unit) for x=0, 0.25, 0.5, 0.75 and 1, respectively. These experimental values straightly follow the half-metallic Slater-Pauling rule as shown in Fig. 3. Based on the first principles calculations, Galanikis *et al.* [16,17] proposed that the $M_s$ of half-metallic Huesler alloys follows the Slater-Pauling rule, $M_s$ =Z-24 ($\mu_B$), where Z is the number of valence electrons. We can thus conclude that these Heusler alloys $Fe_2Co_{1-x}Cr_xSi$ are potential half-metallic candidates.

Very recently, Kokado *et al.* [18] have systematically investigated the sign of the anisotropy magnetoresisetnce (AMR) of ferromagnetic materials. As a result, they found that, for half-metallic ferromagnets, the dominant scattering is s↑→d↑ or s↓→d↓, which cause the sign of the AMR tends to be negative [19,20]. Subsequently, Yang *et al.* [21]measured the AMR ratio in Heusler $Co_2$(Fe,Mn)Si epitaxial films, and pointed out the AMR effect can be an indicator of half-metallicity or non-half-metallicity, which can easily be achieved without having to make any microfabricated device structures. In our work, we also examined the half-metallic properties of $Fe_2Co_{1-x}Cr_xSi$ by AMR effect measurement. The results of normalized in-plane magnetoresistance as a function of magnetic field are displayed in Fig 4. Here a constant current of 5 mA and a magnetic field up to 5 T were applied to the sample plane in the following two configurations: (a) parallel ($\rho_{\parallel}$) and (b) perpendicular ($\rho_{\perp}$) to the current direction, respectively. Besides the MR values, the sign of MR is independent on the configuration between current and magnetic field. The MR of all the samples exhibit no tendency to saturation even at a field of 5T. It is found that the MR sign of $Fe_2CoSi$ is positive and it reaches a value of about 1%. However, the MR sign of other samples is negative. Moreover, the positive MR is almost ten times larger than those with the negative MR sign. Based on the MR measurements, we have calculated the AMR ratio, and the results are shown in Fig 5. Here the AMR ratio was defined as AMR=$[\rho_{\parallel}-\rho_{\perp}]/\rho_{\perp}$. We found that, the AMR ratio of $Fe_2Co_{1-x}Cr_xSi$ (x=0.25, 0.5, 0.75) shows a negative value, which suggests the dominant scattering is s↑→d↑ (see the DOS of $Fe_2Co_{1-x}Cr_xSi$ in Fig. 2). Referring to the AMR description proposed by Kokado *et al.* [18], the negative AMR sign indicates the half-metallic nature of this series. However, for the two end members, $Fe_2CoSi$ and $Fe_2CrSi$, the positive AMR ratio is obtained which suggests the half-metallic properties may be destroyed easily because of their Fermi level located the edge of the gap.

**4. Conclusion**

In summary, using the first principles calculations, we have studied the electronic band structure and the magnetic transport properties of Heusler alloys $Fe_2Co_{1-x}Cr_xSi$. We found that the Fermi level of $Fe_2Co_{1-x}Cr_xSi$ moves from the top of valence band to the bottom of conduction band with increasing Cr content x. a large half-metallic band gap of 0.75 eV is obtained for $Fe_2Co_{0.25}Cr_{0.75}Si$, the Fermi energy is located in the middle point of the minority states with a large ban-gap of 0.75eV. The fabricated Heusler $Fe_2Co_{1-x}Cr_xSi$ ribbon samples are all pure fcc phase with a high degree of order. The saturation magnetic moment of all samples increases lineally with increasing Cr content x, straightly follows the half-metallic Slater-pauling rule. To characterize the half-metallic property, we have further investigated the in-plane anisotropic magnetoresistance (AMR) of the ribbon samples. The negative AMR values are found among the three $Fe_2Co_{1-x}Cr_xSi$ alloys (x=0.25, 0.5, 0.75), meaning they are stable half-metallic ferromagnets.

**Acknowledgements:**

This work was supported by National Natural Science Foundation of China (Grant Nos. 51071172 and 51171207) and National Basic Research Program of China (973 Programs: 2012CB619405).

Figure captions:

FIG. 1. (Color online) Dependence of x-ray diffraction (XRD) patterns on Cr content x for $Fe_2Co_{1-x}Cr_xSi$ (x=0.0, 0.25, 0.5, 0.75, 1). The left inset shows the schematic Heusler structure. The right inset indicates the order parameters as a function of x. Here $s$ and α indicate $B_2$ and $L2_1$ degree of ordering, respectively.

FIG. 2. (Color online) (a) Spin-resolved density of states (DOS) for $Fe_2Co_{1-x}Cr_xSi$ (x=0.0, 0.25, 0.5, 0.75, 1) from first principles calculations. The Fermi level moves from the top of $E_{VB}$ (valence band) to the bottom of $E_{CB}$ (conduction band) with increasing Cr content x. The Cr content x dependence of $E_{CB}$ and $E_{VB}$ (b) and, the sizes of gap $\Delta E_{gap}=E_{VB}-E_{CB}$ (c). The insets in (c) indicate the enlarge profiles for spin-resolved DOS of $Fe_2CoSi$ and $Fe_2Co_{0.25}Cr_{0.75}Si$, respectively.

FIG. 3. (Color online) Saturation magnetization values ($M_s$) for $Fe_2Co_{1-x}Cr_xSi$ (x=0.0, 0.25, 0.5, 0.75, 1) plotted together with the expected $M_s$ from Slater-pauling rule. The inset shows the typical magnetization curves measured by SQUID at 5K.

FIG. 4. (Color online) The normalized in-plane magnetoresistance as a function of magnetic field measured at 10 K for two distinct cases as sketched in the figures: (a) magnetic field parallel to the current and the sample surface and (b) magnetic field parallel to the sample surface but perpendicular to the current.

FIG. 5. (Color online) The AMR ratio with different compositions measured at 10K. Here the AMR ratio was defined as $AMR=[\rho_{\parallel}-\rho_{\perp}]/\rho_{\perp}$

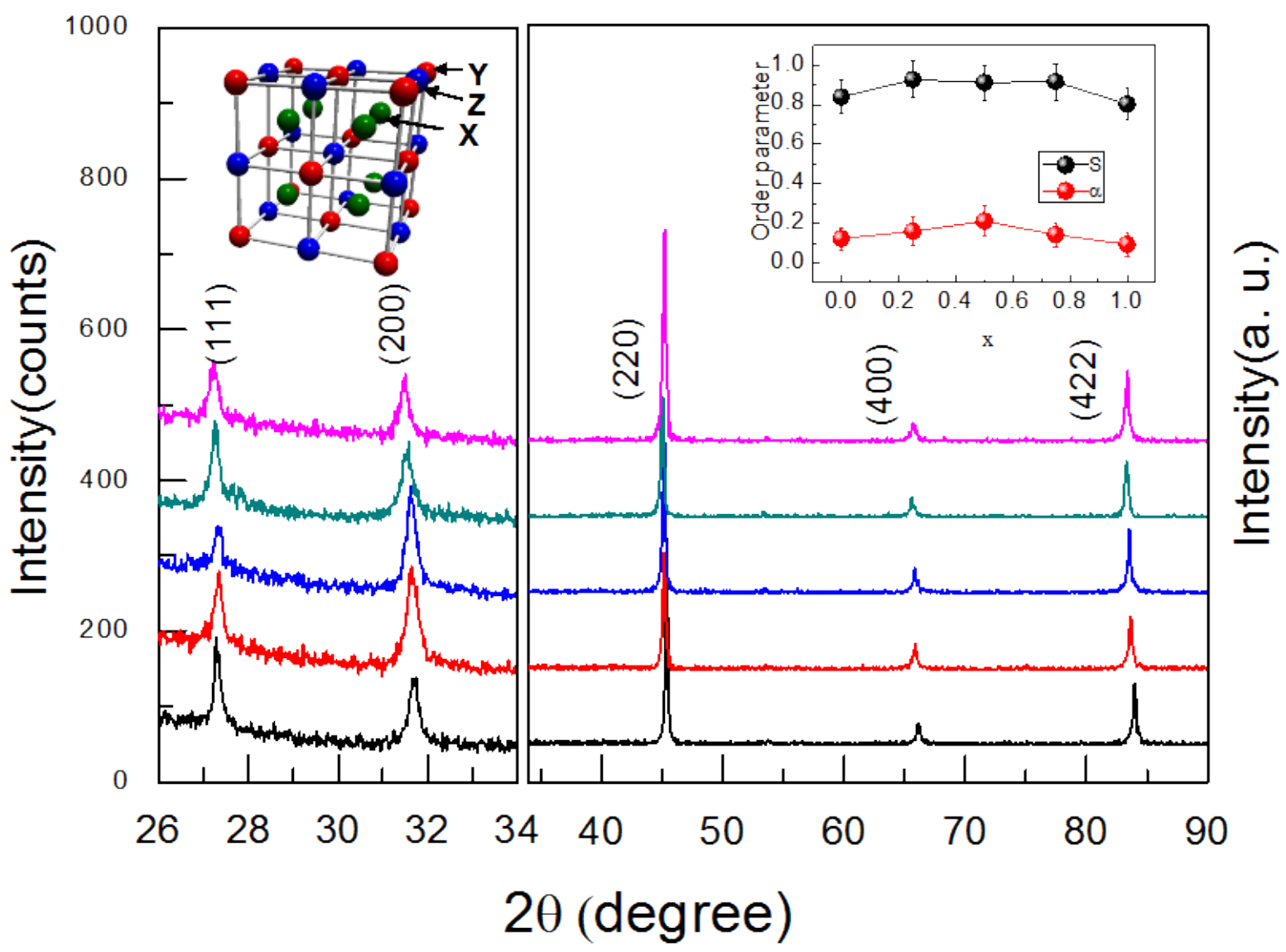

Figure 1, Du et al.,

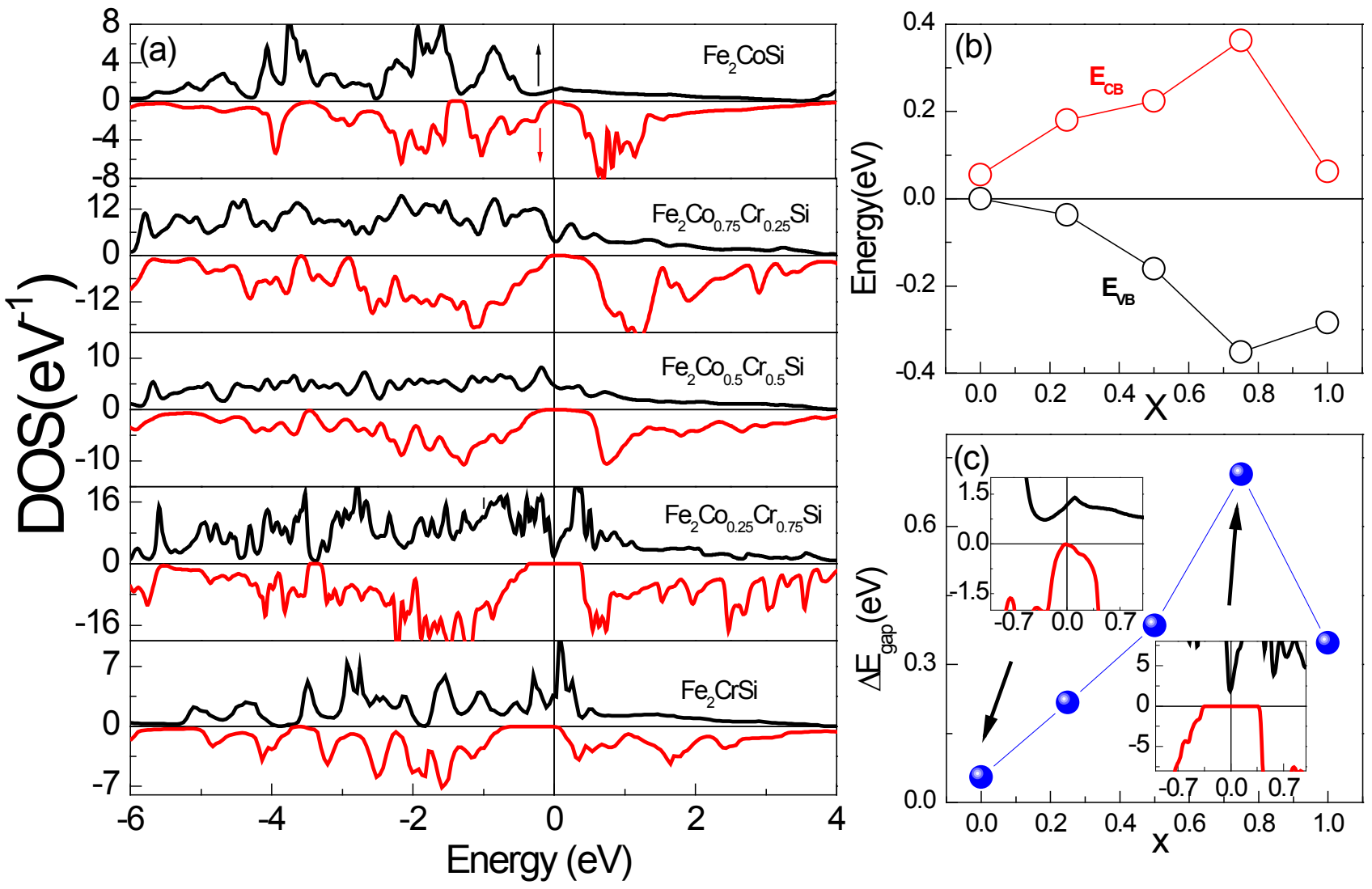

Figure 2, Du et al.,

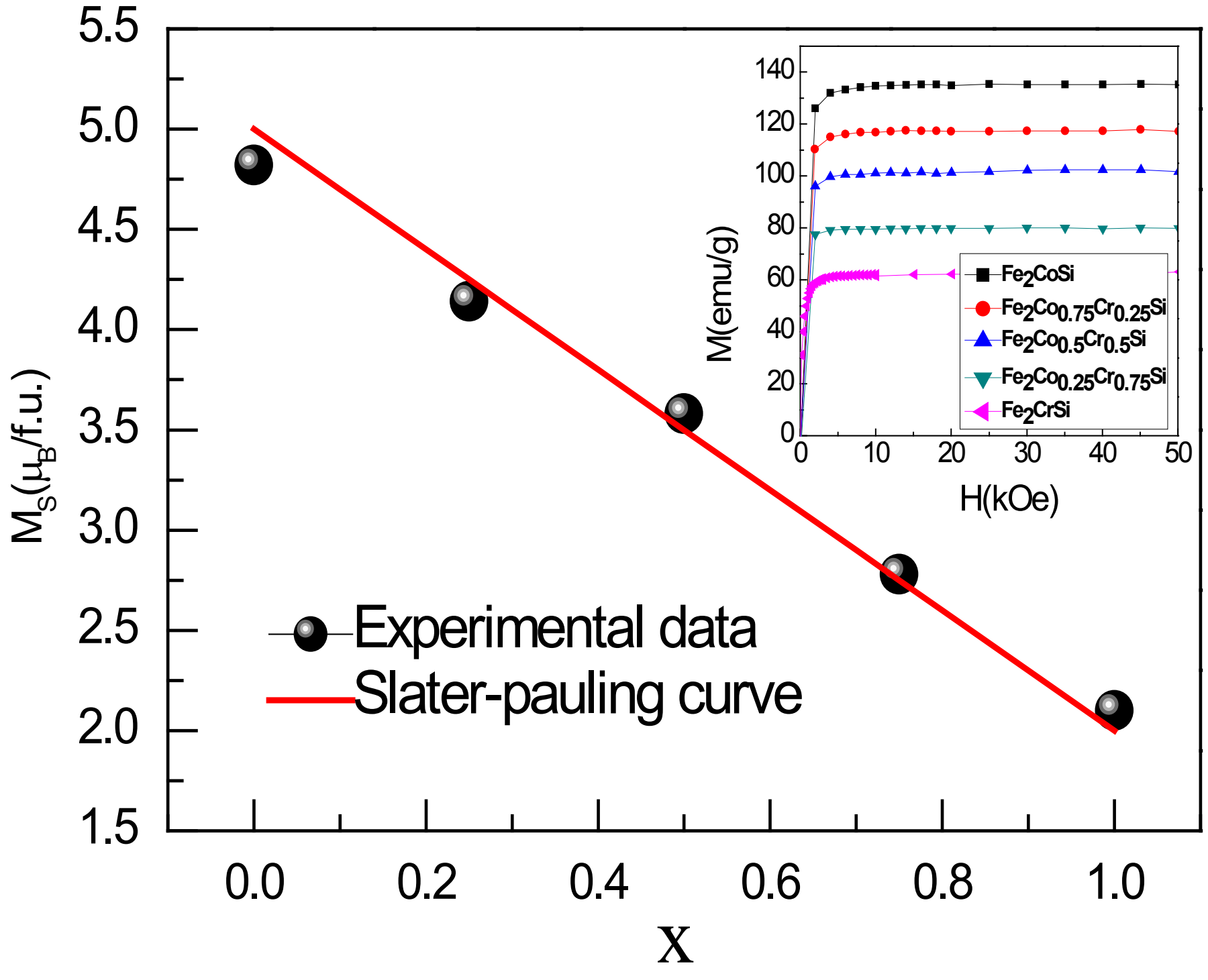

Figure 3, Du et al.,

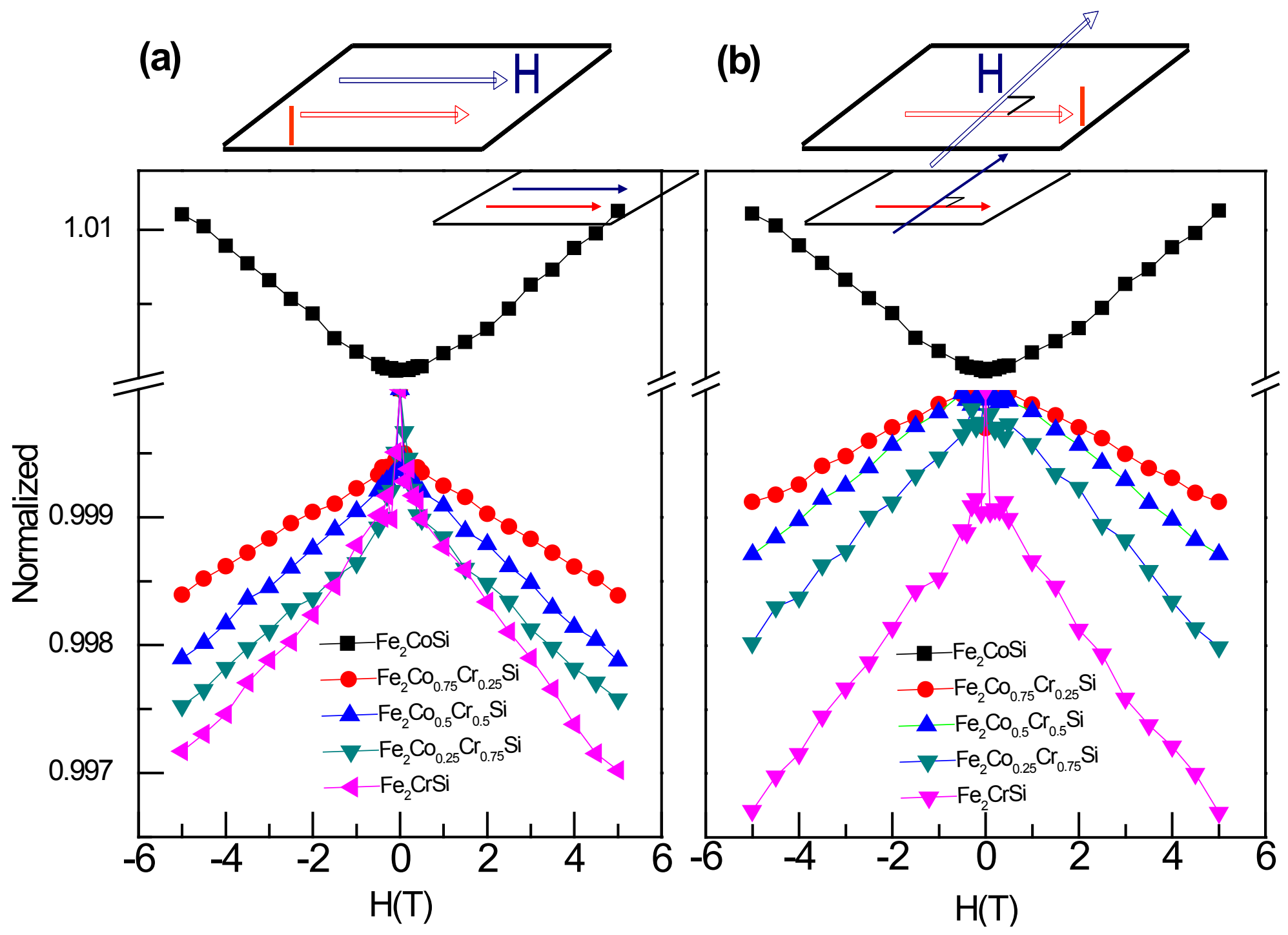

Figure 4, Du et al.,

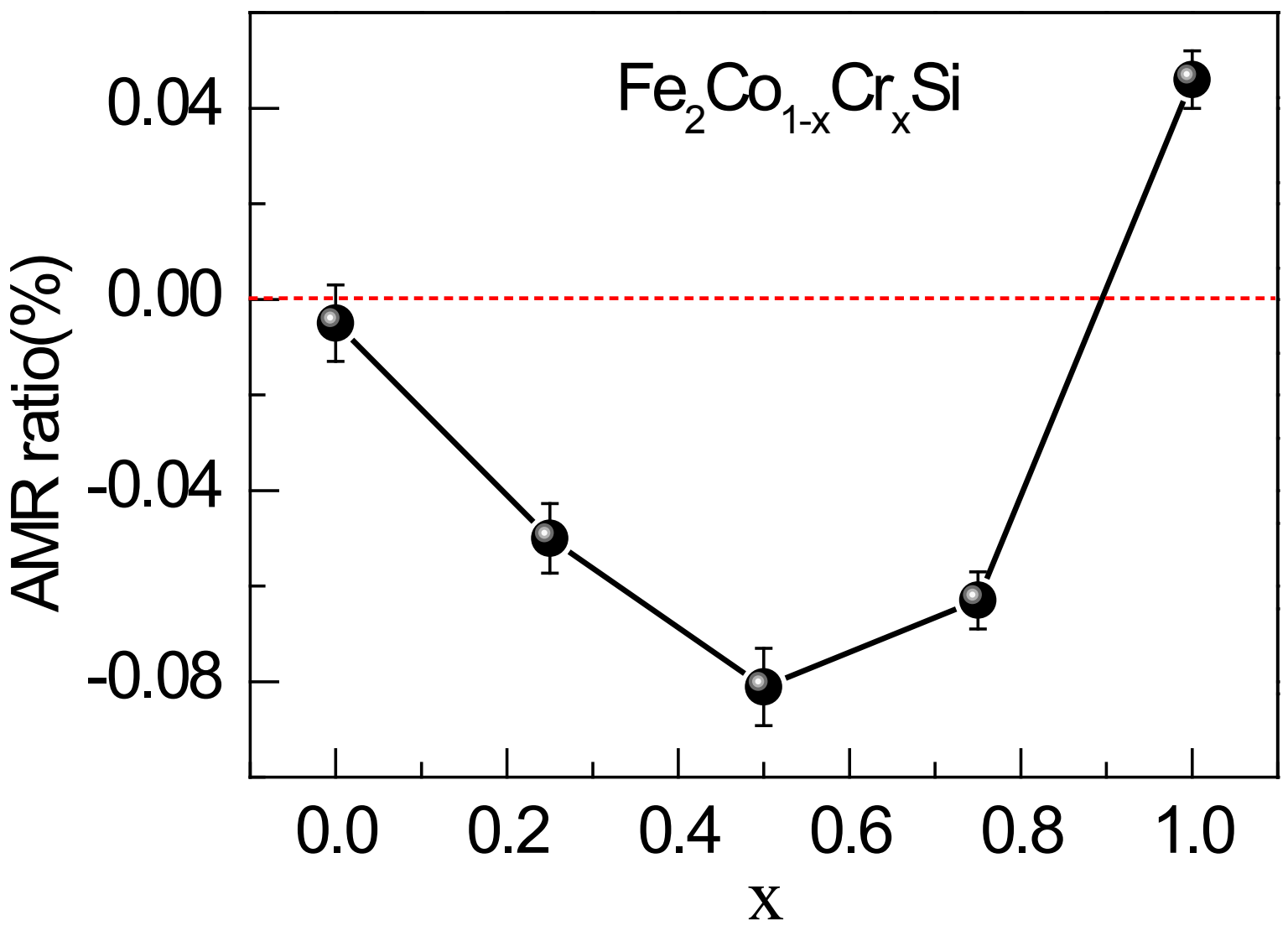

Figure 5, Du et al.,